\RequirePackage{amsthm}
\documentclass[default]{sn-jnl}

\usepackage{graphicx,caption}
\usepackage{multirow}%
\usepackage{amsmath,amssymb,amsfonts}%
\usepackage{amsthm}%
\usepackage{mathrsfs}%
\usepackage[title]{appendix}%
\usepackage{xcolor}%
\usepackage{textcomp}%
\usepackage{manyfoot}%
\usepackage{booktabs}%
\usepackage{algorithm}%
\usepackage{algorithmicx}%
\usepackage{algpseudocode}%
\usepackage{listings}%
\usepackage{url}
\usepackage{hyperref}
\usepackage[encapsulated]{CJK}
\usepackage[numbers]{natbib}
\usepackage{ragged2e}
\usepackage{setspace}
\usepackage{multibib}
\newcites{supp}{Methods References}
\newcommand{\grizli}{\textsc{grizli}}
\newcommand{\arcsec}{\mbox{$^{\prime\prime}$}}

\theoremstyle{thmstyleone}%
\theoremstyle{thmstyletwo}%

\theoremstyle{thmstylethree}%

\newcommand{\msun}{M$_{\odot}$\,}
\newcommand{\Zsun}{\ensuremath{Z_{\odot}}}

\begin{document}

\title[]{Bound star clusters observed in a lensed galaxy 460 Myr after the Big Bang}


\author*[1]{\fnm{Angela} \sur{Adamo}}\email{angela.adamo@astro.su.se}


\author[2]{\fnm{Larry D.} \sur{Bradley}}
\equalcont{These authors contributed equally to this work.}

\author[3]{\fnm{Eros} \sur{Vanzella}}
\equalcont{These authors contributed equally to this work.}

\author[N1]{\fnm{Adélaïde} \sur{Claeyssens}}

\author[4,5]{\fnm{Brian} \sur{Welch}}

\author[6]{\fnm{Jose M} \sur{Diego}}

\author[7,8,9]{\fnm{Guillaume} \sur{Mahler}}

\author[10,11]{\fnm{Masamune} \sur{Oguri}}

\author[12]{\fnm{Keren}\sur{Sharon}}

\author[13,2]{\sur{Abdurro'uf}}

\author[13,2]{\fnm{Tiger Yu-Yang} \sur{Hsiao}}

\author[14, 15]{\fnm{Xinfeng} \sur{Xu}}

\author[3]{\fnm{Matteo} \sur{Messa}}

\author[16,1]{\fnm{Augusto E.} \sur{Lassen}}

\author[17,18]{\fnm{Erik} \sur{Zackrisson}}

\author[19,20]{\fnm{Gabriel} \sur{Brammer}}

\author[2,13,21]{\fnm{Dan} \sur{Coe}}

\author[22]{\fnm{Vasily} \sur{Kokorev}}

\author[4]{\fnm{Massimo} \sur{Ricotti}}

\author[23]{\fnm{Adi} \sur{Zitrin}}

\author[24]{\fnm{Seiji} \sur{Fujimoto}}

\author[25,26]{\fnm{Akio K.} \sur{Inoue}}

\author[13]{\fnm{Tom} \sur{Resseguier}}

\author[5]{Jane R. Rigby}

\author[27,28]{\fnm{Yolanda} \sur{Jim\'enez-Teja}}

\author[29]{Rogier A. Windhorst} 

\author[30,31]{\fnm{Takuya} \sur{Hashimoto}}

\author[32]{\fnm{Yoichi}\sur{Tamura}}

\affil*[1]{\orgdiv{Astronomy Department}, \orgname{Stockholm University \& Oskar Klein Centre}, \orgaddress{\street{Roslagstullsbacken 21}, \city{Stockholm}, \postcode{SE-10691}, \country{Sweden}}}

\affil[2]{\orgname{Space Telescope Science Institute (STScI)}, \orgaddress{\street{3700 San Martin Drive}, \city{Baltimore}, \postcode{21102}, \state{MD}, \country{USA}}}

\affil[3]{\orgdiv{Osservatorio di Astrofisica e Scienza dello Spazio di Bologna}, \orgname{INAF}, \orgaddress{\street{via Gobetti 93/3}, \city{Bologna}, \postcode{40129}, \country{Italy}}}

\affil[4]{\orgdiv{Department of Astronomy}, \orgname{University of Maryland}, \orgaddress{\street{4296 Stadium Drive}, \city{College Park}, \postcode{20742-2421}, \country{USA}}}

\affil[5]{\orgdiv{Astrophysics Science Division, Code 660}, \orgname{NASA Goddard Space Flight Center}, \orgaddress{8800 Greenbelt Rd.}, \city{Greenbelt, MD}, \postcode{20771}, \country{USA}}

\affil[6]{\orgdiv{Instituto de F\'isica de Cantabria }, \orgname{(CSIC-UC)}, \orgaddress{\street{Avda. Los Castros s/n.}, \city{Santander}, \postcode{39005}, \country{Spain}}}

\affil[7]{\orgdiv{STAR Institute},\orgaddress{\street{Quartier Agora - All\'ee du six Ao\^ut, 19c}, \city{Li\`ege}, \postcode{B-4000}, \country{Belgium}}}

\affil[8]{\orgdiv{Centre for Extragalactic Astronomy}, \orgname{Durham University}, \orgaddress{\street{South Road}, \city{Durham},\postcode{DH1 3LE}, \country{UK}}}

\affil[9]{\orgdiv{Institute for Computational Cosmology}, \orgname{Durham University}, \orgaddress{\street{South Road}, \city{Durham},\postcode{DH1 3LE}, \country{UK}}}

\affil[10]{\orgdiv{Center for Frontier Science}, \orgname{Chiba University}, \orgaddress{\street{1-33 Yayoi-cho, Inage-ku}, \city{Chiba}, \postcode{263-8522}, \country{Japan}}}

\affil[11]{\orgdiv{Department of Physics, Graduate School of Science}, \orgname{Chiba University}, \orgaddress{\street{1-33 Yayoi-cho, Inage-ku}, \city{Chiba}, \postcode{263-8522}, \country{Japan}}}

\affil[12]{Department of Astronomy, University of Michigan, 1085 S. University Ave, Ann Arbor, MI 48109, USA}

\affil[13]{\orgdiv{Center for Astrophysical Sciences, Department of Physics and Astronomy}, \orgname{The Johns Hopkins University}, \orgaddress{\street{3400 N Charles St.}, \city{Baltimore}, \postcode{MD 21218}, \country{USA}}}

\affil[14]{\orgdiv{Department of Physics and Astronomy}, \orgname{Northwestern University}, \orgaddress{\street{2145 Sheridan Road}, \city{Evanston, IL}, \postcode{60208}, \country{USA}}}

\affil[15]{\orgdiv{Center for Interdisciplinary Exploration and Research in Astrophysics (CIERA)}, \orgname{Northwestern University}, \orgaddress{\street{1800 Sherman Avenue}, \city{Evanston, IL}, \postcode{60201}, \country{USA}}}

\affil[16]{\orgdiv{Instituto de Física, Departamento de Astronomia}, \orgname{Universe Federal do Rio Grande do Sul}, \orgaddress{\street{Avenida Bento Gonçalves, 9500}, \city{Porto Alegre}, \postcode{91509-900}, \country{Brazil}}}

\affil[17]{\orgdiv{Observational Astrophysics, Department of Physics and Astronomy}, \orgname{Uppsala University}, \orgaddress{\street{Box 516}, \city{Uppsala}, \postcode{SE-751 20}, \country{Sweden}}}

\affil[18]{\orgname{Swedish Collegium for Advanced Study}, \orgaddress{\street{Linneanum, Thunbergsvägen 2}, \city{Uppsala}, \postcode{SE-752 38 Uppsala}, \country{Sweden}}}

\affil[19]{\orgname{Cosmic Dawn Center (DAWN)}, \orgaddress{\city{Copenhagen}, \country{Denmark}}}

\affil[20]{\orgname{Niels Bohr Institute, University of Copenhagen}, \orgaddress{\street{Jagtvej 128}, \city{Copenhagen}, \country{Denmark}}}

\affil[21]{Association of Universities for Research in Astronomy (AURA) for the European Space Agency (ESA), STScI, Baltimore, MD, USA}

\affil[22]{\orgdiv{Kapteyn Astronomical Institute}, \orgname{University of Groningen}, \orgaddress{\street{Landleven 12}, \city{Groningen}, \postcode{9700 AV}, \country{Netherlands}}}

\affil[23]{\orgdiv{Department of Physics}, \orgname{Ben-Gurion University of the Negev}, \orgaddress{\street{P.O. Box 653}, \city{Be'er-Sheva}, \postcode{84105}, \country{Israel}}}

\affil[24]{Department of Astronomy, The University of Texas at Austin, Austin, TX 78712, USA}

\affil[25]{\orgdiv{Department of Physics, School of Advanced Science and Engineering, Faculty of Science and Engineering}, \orgname{Waseda University}, \orgaddress{\street{3-4-1 Okubo}, \city{Shinjuku, Tokyo}, \postcode{169-8555}, \country{Japan}}}

\affil[26]{\orgdiv{Waseda Research Institute for Science and Engineering, Faculty of Science and Engineering}, \orgname{Waseda University}, \orgaddress{\street{3-4-1 Okubo}, \city{Shinjuku, Tokyo}, \postcode{169-8555}, \country{Japan}}}

\affil[27]{\orgdiv{Instituto de Astrof\'isica de Andaluc\'ia}, \orgname{(CSIC)}, \orgaddress{\street{Glorieta de la Astronom\'ia s/n.}, \city{Granada}, \postcode{18008}, \country{Spain}}}

\affil[28]{\orgdiv{Observat\'orio Nacional}, \orgname{(MCTI)}, \orgaddress{\street{Rua Gal. Jos\'e Cristino 77, S\~{a}o Crist\'ov\~{a}o}, \city{Rio de Janeiro}, \postcode{20921-400}, \country{Brazil}}}

\affil[29]{School of Earth and Space Exploration, Arizona State University, Tempe, AZ 85287-1404, USA}

\affil[30]{\orgdiv{Division of Physics, Faculty of Pure and Applied Sciences}, \orgname{University of Tsukuba}, \orgaddress{\street{1-1-1 Tennodai}, \city{Tsukuba, Ibaraki}, \postcode{305-8571}, \country{Japan}}}

\affil[31]{\orgdiv{Tomonaga Center for the History of the Universe (TCHoU)}, \orgname{University of Tsukuba}, \orgaddress{\street{1-1-1 Tennodai}, \city{Tsukuba, Ibaraki}, \postcode{305-8571}, \country{Japan}}}

\affil[32]{\orgdiv{Department of Physics, Graduate School of Science}, \orgname{Nagoya University}, \orgaddress{\street{Furo, Chikusa}, \city{Nagoya}, \postcode{464-8602}, \country{Japan}}}


\maketitle

\textbf{The Cosmic Gems arc is among the brightest and highly magnified galaxies observed at redshift $z\sim10.2$ \citep{bradley2024}. However, it is an intrinsically UV faint galaxy, in the range of those now thought to drive the reionization of the universe \citep{roberts-borsani2023, atek2024, simmonds2024MNRAS.527.6139S}. Hitherto the smallest features resolved in a galaxy at a comparable redshift are between a few hundreds and a few tens of parsecs \citep{hsiao2023, morishita2024}. Here we report JWST observations of the Cosmic Gems. The light of the galaxy is resolved into five star clusters located in a region smaller than 70 parsec. They exhibit minimal dust attenuation and low metallicity, ages younger than 50 Myr and intrinsic masses of $\sim10^6$ \msun. Their lensing-corrected sizes are approximately 1 pc, resulting in stellar surface densities near $10^5$~\msun/pc$^2$, three orders of magnitude higher than typical young star clusters in the local universe \citep{brown2021}.  
Despite the uncertainties inherent to the lensing model, they are consistent with being gravitationally bound stellar systems, i.e., proto-globular clusters. We conclude that star cluster formation and feedback likely contributed to shape the properties of galaxies during the epoch of reionization.} \\

The Cosmic Gems arc (SPT0615-JD1) was initially discovered in HST images obtained by the RELICS survey of the lensing galaxy cluster SPT-CL~J0615$-$5746 at $z = 0.972$ and reported as a redshift $z=10$ candidate \citep{Salmon2018}.

\begin{figure}[ht]%
\centering
\includegraphics[width=1.0\textwidth]{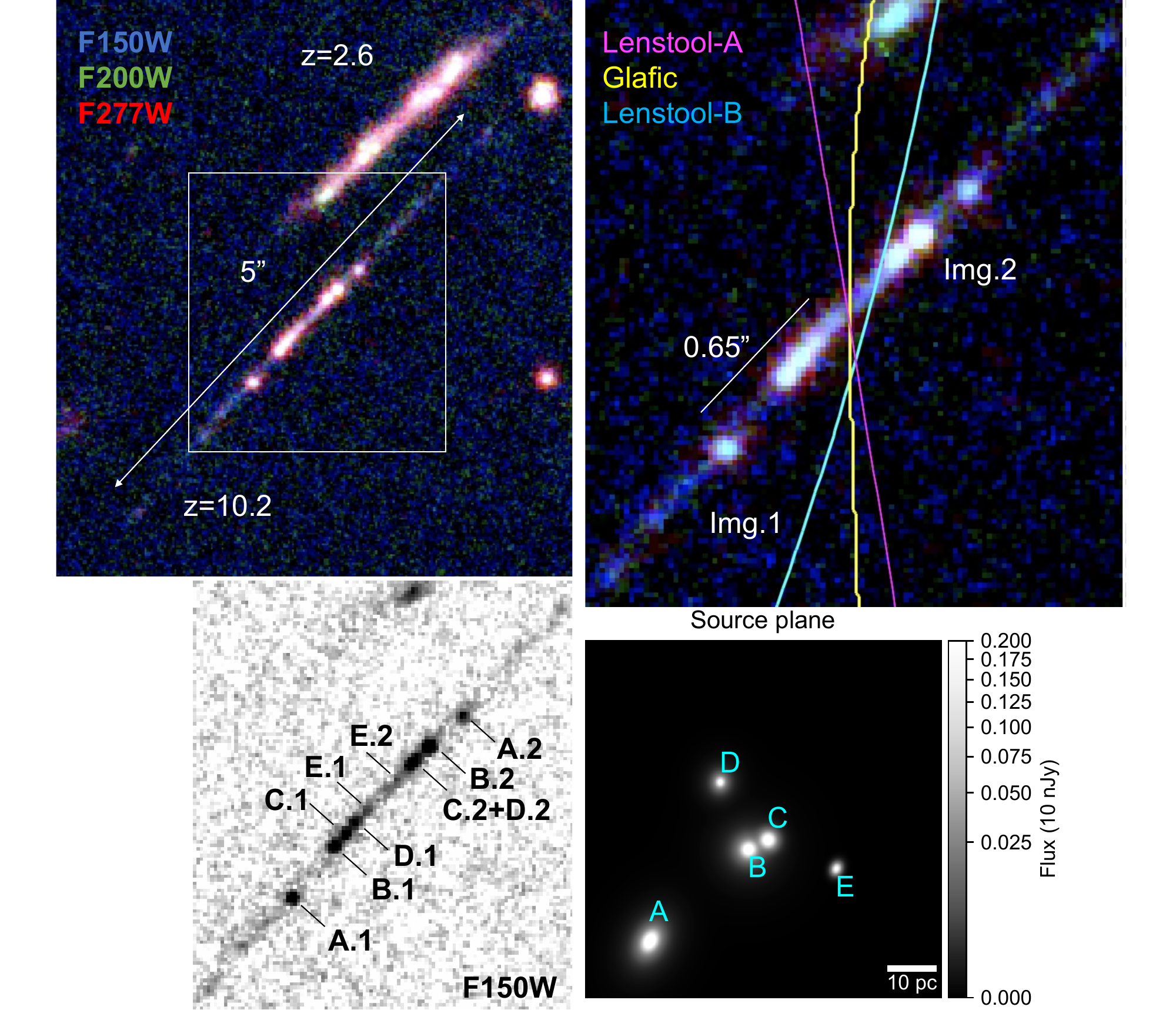}
\caption{{\bf The Cosmic Gems arc in a JWST color-composite.} The filter combination shows the rest-frame UV, blue optical wavelengths (1200--2800~\AA). The arc is extended over 5$\arcsec$. A foreground galaxy at redshift $z_{\rm phot}=2.6$ is visible above and to the right. The field of view is rotated North up. Top right: A zoom inset of the center of the arc where the brightest star clusters are located highlighted by the white square on the left image. Two mirror images are observed due to gravitational lensing. 
Lensing critical curves based on three models are shown bisecting the arc.
Bottom left: each star cluster is labeled in a grayscale FUV rest-frame image of the galaxy. Bottom right: Source plane reconstruction of the core of the galaxy where the star clusters are located showing their relative sizes and positions. The physical distance between A and E and A and D is about 40 pc. Note there is some uncertainty in source positions parallel to the lensing caustic.}
\label{fig1}
\end{figure}

A recent JWST/NIRCam imaging campaign of this field has observed the Cosmic Gems 
arc with eight bands covering 0.8 to 5 $\mu$m range (Methods). 
Spectral energy distribution (SED) fitting to the JWST photometry indicates that the Cosmic Gems galaxy has a fairly young stellar population with mass-weighted age $<79$~Myr and a lensing-corrected stellar mass in the range 2.4--5.6 $\times 10^{7}$~\msun, with low dust extinction ($A_V<0.15$ mag) and metallicity ($<1$\% \Zsun) \citep{bradley2024}. 

The FUV--to--optical rest-frame of Cosmic Gems arc 
reveals bright clumpy structures 
and extended faint emission over a $5''$-long arc (Fig.~1). The symmetry between the south--east (hereafter \emph{Img.1}) and the north--west (\emph{Img.2}) part of the arc uncovers two lensed mirror images of the galaxy, implying that the Cosmic Gems arc is observed at very high magnification on the lensing critical curve. Four independent magnification models have been created to account for the galaxy appearance. All the models successfully reproduce the $z=10.2$ critical line crossing the Cosmic Gems arc (Methods). 

Five star cluster candidates are uniquely identified in \emph{Img.1}. In \emph{Img.2}, three sources are clearly distinguishable in the F150W filter (Fig.~1), along with a fourth fainter source (E.2). The appearance of \emph{Img.2} is likely perturbed by further lensing effects due to the northern galaxy at $z\sim2.6$ (visible in the upper right corner) and possibly by an undetected small scale perturber closer to the arc \citep{bradley2024}. Source D.2 is possibly blended with C.2, and is therefore identified for the remaining of the analysis as C.2$+$D.2. The source E.2 is only detected at a 2$\sigma$ level and not included in this analysis (Methods). 
The observed projected distance between the A.1 and E.1 clusters in \emph{Img.1} is about $0.65''$. Using forward modelling and \emph{Lenstool-A} predictions presented in Methods, we find that their physical distance is $42_{-5}^{+29}$ pc. The star clusters are all located within this compact region (Fig.~1). 
 The size of this region is similar to that reported for individual stellar clumps observed in the moderately-lensed galaxies at redshift $z \sim 10$ \citep{hsiao2023, roberts-borsani2023}. 

\begin{table}[ht]
\caption{{\bf Estimated physical properties of the Cosmic Gems arc star clusters.} We report de-convolved observed half-light radii in pixels, R$_{\rm eff, obs}$, and lensing-corrected R$_{\rm eff}$ and median stellar masses M$_{*, int}$ using magnifications produced by the reference \emph{Lenstool-A} model. M$_{*}$ are recovered from the \textsc{BAGPIPES-}\emph{exp} reference fit. Errors are estimated from 68 \% confidence level of the distributions. These quantities have been used to determine stellar surface density, $\Sigma_{*}$, and dynamical age $\Pi$ listed in the last columns. Evaluation of magnification uncertainties are discussed in Methods.} 
\label{res:tab}%
\begin{tabular}{@{}ccccccc@{}}
\toprule
ID & R$_{\rm eff, obs}$ & R$_{\rm eff}$ &R$_{\rm eff, FM}$ &M$_{*, int}$ & $\Sigma_{*}$ & $\log(\Pi)$ \\
 &  [px] &  [pc] & [pc] & [$10^6$\msun] &  [$10^5$\msun/pc$^2$] &\\
\midrule
A.1 & $ 0.6 ^{+ 0.4 }_{- 0.1 }$ & $ 1.1 ^{+ 0.7 }_{- 0.2 }$ &$1.1 \pm 0.1$&$ 2.45 ^{+ 5.20 }_{- 1.56 }$&$ 1.92 ^{+ 1.60 }_{- 1.44 }$&$ 1.94 ^{+ 0.71 }_{- 0.27 }$\\
B.1 & $ 1.1 ^{+ 0.1 }_{- 0.5 }$ & $ 1.1 ^{+ 0.1 }_{- 0.5 }$ &$0.9 \pm 0.1$&$2.65 ^{+ 1.09 }_{- 1.26 }$&$ 1.93 ^{+ 4.16 }_{- 1.11 }$&$ 2.11 ^{+ 0.83 }_{- 0.50 }$\\
C.1 & $ <1.25$ &$ <1 $ &$0.9 \pm 0.2$& $1.13 ^{+ 1.77 }_{- 0.65 }$& $>1.3$&$ >1.90$\\
D.1  & $ 1.2 ^{+ 0.2 }_{- 1.1 }$ &$ 0.6 ^{+ 0.1 }_{- 0.6 }$ & $0.8 \pm 0.2$&$1.13 ^{+ 1.23 }_{- 0.74 }$&$ 2.39 ^{+ 7.41 }_{- 1.98 }$&$ 2.17 ^{+ 0.85 }_{- 1.03 }$\\
E.1  & $ 1.5 ^{+ 0.7 }_{- 0.5 }$ &$ 0.4 ^{+ 0.2 }_{- 0.1 }$ & $0.7 \pm 0.2$& $ 1.01 ^{+ 0.37 }_{- 0.36 }$&$ 6.92 ^{+ 4.90 }_{- 4.22 }$&$ 3.06 ^{+ 0.32 }_{- 0.59 }$\\
A.2  & $ 1.0 ^{+ 0.4 }_{- 0.3 }$ & $ 1.7 ^{+ 0.8 }_{- 0.4 }$&$1.3 \pm 0.1$& $2.89 ^{+ 1.56 }_{- 1.35 }$ &$ 0.88 ^{+ 0.98 }_{- 0.46 }$&$ 1.94 ^{+ 0.52 }_{- 0.41 }$\\
B.2  & $ <1.25$ &$ <1.4 $ &$1.0 \pm 0.04$& $3.01 ^{+ 3.21 }_{- 1.61 }$& $>5.10$& $>1.8$\\
C.2$+$D.2 & $ 2.7 ^{+ 16.1 }_{- 2.6 }$ & $ 1.9 ^{+ 11.4 }_{- 1.9 }$ & $0.9 \pm 0.1$& $4.36 ^{+ 0.98 }_{- 1.90 }$&$ 1.05 ^{+ 7.89 }_{- 0.89 }$&$ 1.99 ^{+ 1.12 }_{- 1.49 }$\\
\botrule
\end{tabular}
\end{table}

The intrinsic physical properties of these 5 star clusters are particularly meaningful for probing proto-GC formation mechanisms as well as their potential evolution. As described in Methods, we find that A.1 is only marginally resolved, with an observed effective half-light radius $R_{\rm eff,obs}= 0.6$ px, while C.1 and B.2 are consistent with being unresolved. For the latter  sources, we assumed an upper-limit to their radii coincident with the half-width half maximum (HWHM) of the stellar PSF in the F150W ($0.025''= 1.25$ px). To derive lensing-corrected R$_{\rm eff}$, we assumed the predicted \emph{Lenstool-A} tangential magnifications at the location of the star clusters (listed in Extended Data Table~3). The 5 star clusters have intrinsic R$_{\rm eff}$ close to 1 pc (within uncertainties, Table~1). Using  the other lensing models produces similar size ranges (0.3 to 0.9 pc for \emph{Lenstool-B} and 0.3 to 1.2 pc for \emph{Glafic}). Independent intrinsic sizes (R$_{\rm eff, FM}$ in Table~1) have been derived by projecting the star cluster shapes from the source plane into the image plane. The latter method recovered intrinsic sizes in excellent agreement, within the uncertainties, with those measured in the image plane, strengthening the reliability of the derived values. 

The star clusters have been fitted with BAGPIPES \citep{carnall2018} and PROSPECTOR \citep{johnson2021}. We tested different star formation history (SFH) assumptions that simulate a single burst (inherent to the small sizes of the stellar systems analysed), different high-mass limits of the initial mass function (IMF), and models with stellar binaries (presented in Methods). Despite the assumptions, the resulting cluster physical properties (ages, masses, extinction, metallicities) are in reasonable agreement. In the analysis presented here, we use the physical values derived with a SFH based on a single exponential decline with $\tau=1$ Myr (listed in Extended Data Table~2).

The recovered ages of the star cluster candidates are between 9 and 36 Myr.
The age range suggests that star formation has been propagating within this compact area of the galaxy for a few tens of Myr. 
The measured rest-frame UV slopes of the star clusters ($\beta$ between $-1.8$ and $-2.5$, with $F_{\lambda} \sim \lambda^{\beta}$, listed in Extended Data Table~1) 
are similar to those found for more evolved star clusters in the Sunburst arc at redshift 2.37 \citep{kim2023}. While the Cosmic Gems clusters are not extremely young, they have likely delivered large amounts of energy and momentum to their host galaxy. 

The lensing-corrected stellar masses range between 1.0 -- 2.6 $\times$ $10^6$ \msun, for a total combined stellar mass of $8.3\times10^6$ \msun. The total mass of the clusters is close to 30\% of the total stellar mass of the host. Since the mass-weighted age of the galaxy and those of the star clusters are comparable, one can extrapolate the cluster formation efficiency (CFE) \citep{adamo2020} to be around 30 \%. 
A caution note is necessary, since the mass estimates (both for the galaxy and star clusters) are subjected to magnification values and SED fit uncertainties, making the quoted CFE uncertain. 
A more direct way to establish the CFE is to use the fraction of observed FUV light in star clusters with respect to the host. This quantity is not affected by the same degeneracy as the mass estimates and, thus, is a more reliable indicator of the CFE, under the assumption that the FUV light is produced by stellar populations formed during a similar timescale (as we find here). The analysed star clusters account for $\sim$ 60 \% of the total F150W flux  of the host  extracted within an elliptical Kron aperture (0.51$\pm0.01$ $\mu$Jy, corresponding to an intrinsic FUV ABmag of $-17.8$ after lensing correction, \citep{bradley2024}), thus reinforcing the conclusion that star formation in star clusters is a major mode for the Cosmic Gems arc and  high-redshift galaxies with similar physical properties.  This observationally driven conclusion is supported by high-resolution numerical simulations \citep{Garcia2023} and analytical models \citep{nebrin2023} that find that compact star clusters with sizes 0.5 -- 2~pc are the dominant star formation mode in the first low-metallicity dwarf galaxies. The compactness of the star clusters appears also to drive the leakage of hydrogen ionizing radiation from their natal molecular cloud  \citep{He2020}, making the star clusters observed here potential contributors to cosmic reionization. Massive star clusters like those observed in the Cosmic Gems arc are predicted by the feedback-free starburst model by \citep{dekel2023}, and could be at the root of the super-Eddington conditions necessary to launch strong outflows in short timescales \citep{ferrara2023}, both models aimed to explain the bright UV luminosity reported for $z>9$ galaxies.

\begin{figure}[ht]%
\centering
\includegraphics[width=0.6\textwidth]{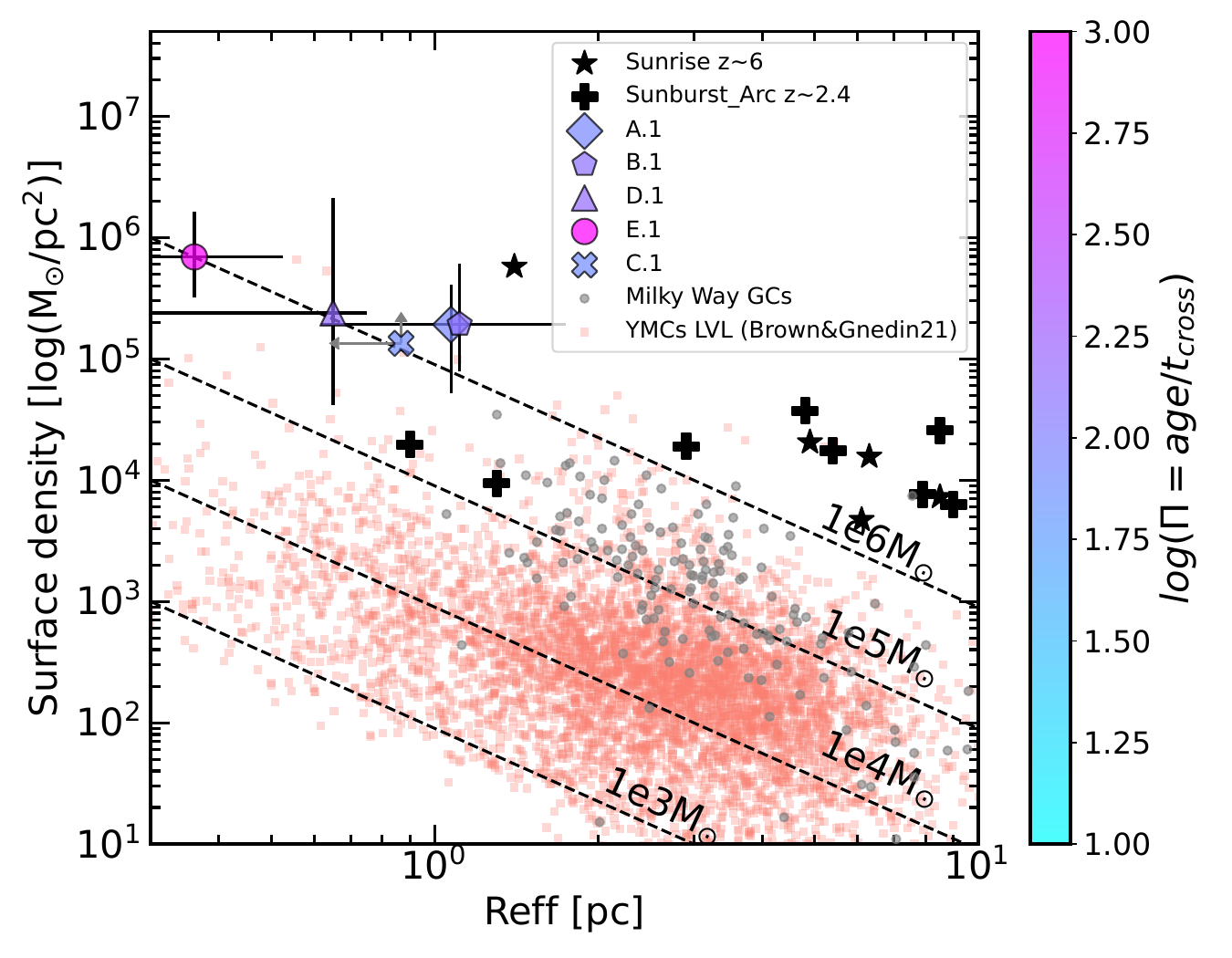}
\caption{{\bf Cluster stellar surface density versus half-light radius R$_{\rm eff}$.} The star clusters in the $z_{\rm phot} \sim 10.2$ Cosmic Gems arc are shown color-coded by their dynamical age, $\log(\Pi)$. The plotted values have been derived using the reference lensing model. Predictions by other models do not change the observed trends. The error bars do not account for magnification uncertainties. However, we refer the reader to Extended Data Fig.~4, where we show that magnification uncertainties do not affect these results. Other gravitationally-bound star clusters detected in lensed galaxies at $z \sim 6$ and $z = 2.38$ are included, along with the $z = 0$ Milky Way globular clusters and young star clusters properties of star-forming spiral galaxies in the Local Volume (distance$<$16 Mpc). Lines of equal mass show the change in density as a function of R$_{\rm eff}$.}
\label{fig2}
\end{figure}

The resulting stellar surface densities of the clusters are $\sim 10^5$ \msun/pc$^2$ (Table~\ref{res:tab}). Consistent physical properties have been reported in star clusters detected in the Sunrise arc at $z \sim 6$ \citep{vanzella2023} and the Sunburst arc at $z = 2.37$ \citep{vanzella2022} (see Fig.~2). Using the derived ages, masses, and intrinsic sizes, we also determine whether these stellar systems are gravitationally bound. According to the framework introduced by \cite{gieles2011}, a star cluster
is considered bound if its age is greater
than the crossing time of the system (where $t_{\rm cross} = 10 \sqrt{R_{\rm eff}^3/GM}$), or in other words, under the assumption of virial equilibrium, a cluster is gravitationally bound if $\Pi = {\rm Age}/{\rm t_{cross}} > 1$. The $\log(\Pi)$ values reported in Table~\ref{res:tab} are all significantly larger than unity,

indicating that we are indeed detecting star clusters in an early galaxy, 460 Myr after the Big Bang. This conclusion is valid in spite of the uncertainties inherent to physical quantity estimates as well as lensing models.

The Cosmic Gems arc clusters (Fig.~2) have significantly higher stellar densities and smaller sizes than typical young star clusters observed in the local universe \citep{brown2021} as well as GCs in the Milky Way \citep{Baumgardt2023}. The offset with respect to young star clusters in the local universe is expected since the conditions under which star formation operates in reionization--era galaxies are more extreme (e.g, more compact, harbour harder ionising radiation fields, and reach higher electron densities and temperatures,\citep{morishita2024,roberts-borsani2024}). The offset with respect to local GCs could be explained in terms of dynamical evolution. GCs are hot stellar systems where stars continuously exchange energy and momentum. Three different internal mechanisms contribute to their dynamical evolution over a Hubble time: (i) mass-loss due to stellar evolution; (ii) relaxation due to N-body interactions; (iii) formation and dynamics of stellar black holes (SBHs) \citep{gieles2010,antonini2020}. Mass loss due to stellar evolution drives GCs' adiabatic expansion under the condition of virial equilibrium. A typical mass-loss of 50 \% will expand the initial radius of the cosmic Gems proto-GCs by a factor of 2, while density will decrease correspondingly by a factor of 8 \citep{gieles2010}. Relaxation time scales (shortened by the presence of SBHs \citep{wang2020}) will also contribute to their expansion. Finally, external tidal fields will further affect the dynamical evolution of these bound stellar systems, which appear to be \emph{bona fide} proto-GCs.

Very dense stellar clusters ($\Sigma_* \sim 10^5$ \msun/pc$^2$, which for R$_{\rm eff}$ = 1 pc correspond to $\rho_h\sim 10^5$ \msun/pc$^3$), like those detected in the Cosmic Gems arc, are predicted to form in low metallicity and highly dense gas \citep{fukushima2023}, where radiative pressure cannot counteract the collapse, resulting in extremely high star formation efficiencies ($\sim$80\% \citep{menon2023}). The high stellar densities found in these proto-GCs imply a significant increase in stellar BH mergers in their interiors \citep{choksi2019, antonini2023} and therefore pave the way to intermediate mass BH seeds \citep{Katz2015}. With stellar masses $>10^5$ \msun, these star clusters naturally harbour Wolf-Rayet and very massive stars \citep{crowther2016}, and because of their  elevated stellar densities, satisfy the necessary condition to form supermassive stars in runaway collisions within their cores \citep{charbonnel2023}. These different classes of stars are among the potential polluters that could explain the observed nitrogen enrichment in the ionised gas of high-redshift galaxies \citep{marques2023}, possibly linked to the formation of the chemically enriched stellar populations ubiquitously found in Milky Way GCs \citep{gratton2019}. 

Cosmological simulations that focus on Milky Way disk-like assembly find that the majority of its GC population form at redshift~$z < 7$ \citep{reina-campos2019, grudic2023, belokurov2023}, suggesting that these star clusters forming at z$\sim$10 might build up the GC populations of more massive early type galaxies in the local Universe. It is difficult to predict whether the proto-GCs of the Cosmic Gems arc will survive a Hubble time. Their chances would be highly enhanced if they were ejected into their host halo during dynamical interactions (e.g., \citep{adamo2020}).

\section*{Methods}\label{methods}

\subsection*{Size \& flux measurements}

The JWST NIRCam \citep{NIRCAM2023} observations of the SPT-CL~J0615--5746 galaxy cluster were obtained in 2023 September (GO 4212: PI Bradley) using four short-wavelength (SW) filters (F090W, F115W, F150W, and F200W) and four long-wavelength (LW) filters (F277W, F356W, F410M, and F444W) spanning 0.8 -- 5.0~$\mu m$.  Each filter had an exposure time of 2920.4~s. The data were reduced with the \grizli\ (version 1.9.5) reduction package \citep{Grizli}.  They suffered from strong wisps \citep{rigby2023} and required special background subtraction as described in \citep{bradley2024}. The final data are in units of 10 nJy. The NIRCam SW (LW) images were drizzled to a pixel scale of 0.02 (0.04)\arcsec/pixel. For further details on the observations and image reduction, please see \citep{bradley2024}. We assume throughout the analysis a cosmology with $H_0=67.7$ km~s$^{-1}$~Mpc$^{-1}$ and $\Omega_m=0.31$ \citep{Planck18_cosmo}. Under these assumptions, 1 SW pixel (0.02\arcsec) corresponds to 83.7 pc at $z=10.2$.

\begin{figure}[ht]%
\centering
\includegraphics[width=0.6\textwidth]{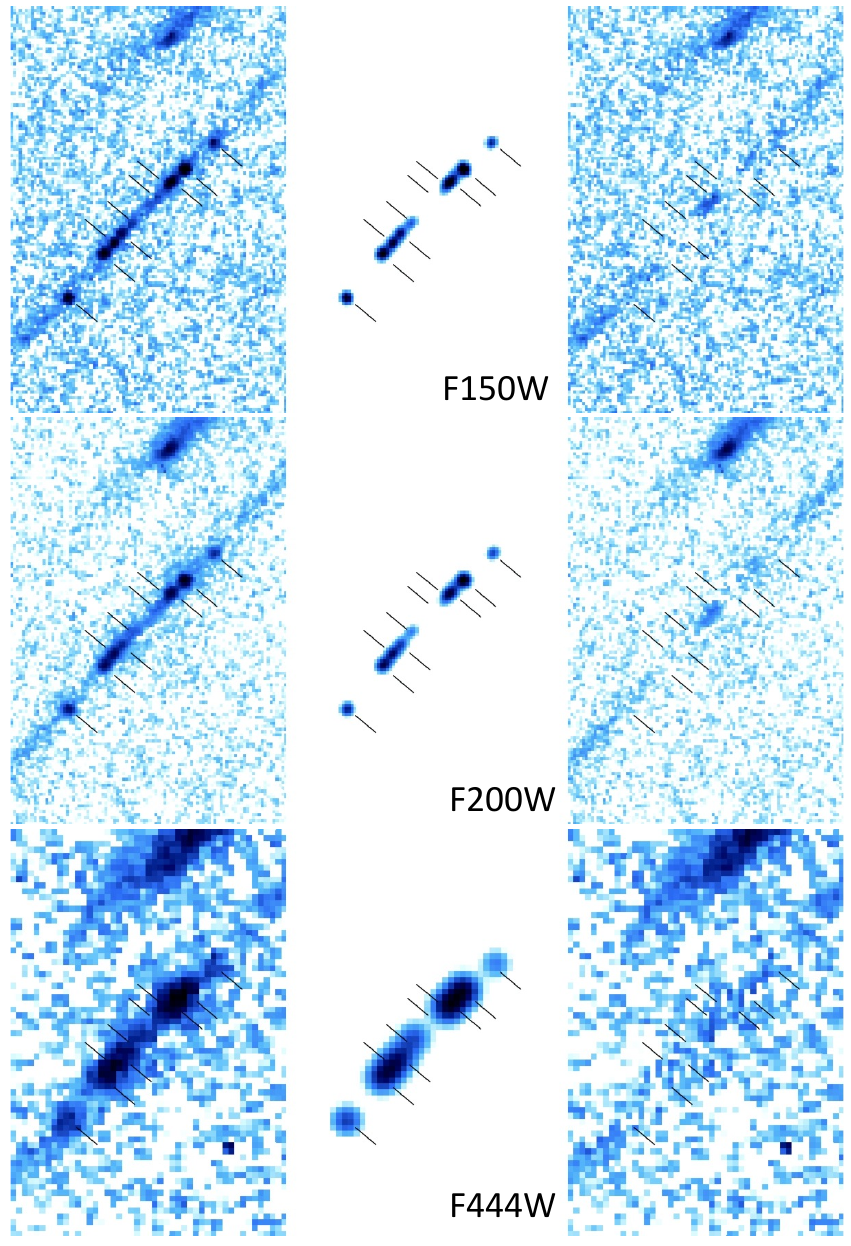}
\caption*{{\bf Extended Data Figure 1}. {\bf Cluster light fit in the image plane}. Observed JWST images (left), best-fitted clump shape after removing the local diffuse light, (centre), and residual images (right) in the reference filter F150W (top), and two more bands, the F200W (middle) with similar resolution to the F150W, and the F444W (bottom), with the lowest spatial resolution. We show log-scale images matched in flux in each band. } 
\end{figure}

We derived the star cluster radii and multiband photometry by applying the method published and tested in \cite{messa2019, messa2022, claeyssens2023}. We simultaneously fitted for the shape of the light distribution, the flux and the local background (including the galaxy diffuse light) of each of the identified clusters in the reference filter, F150W, which offers the sharpest view of the clusters. Empirical PSFs in all filters used in this work were built by selecting stars in each band. We fitted the PSF out to 0.4\arcsec\, with an analytical expression, which was then convolved with varying 2D Gaussians to create a grid of models. Intrinsic sizes (de-convolved by PSF) have then been derived by fitting the observed light distribution of each cluster with this grid of 2D Gaussian models in the F150W.  
Due to the large shear effects, compact sources might appear resolved in the shear direction ($y_{\rm std}$) \citep[e.g.,][]{vanzella2019}. We assumed that the measured major-axis y$_{\rm std}$ of the 2D Gaussian ellipse is the standard deviation of a 2D circular Gaussian which we translated into the observed PSF-deconvolved effective radius, R$_{\rm eff,obs}=y_{\rm std}\times \sqrt{2 ln(2)}$. The derived R$_{\rm eff,obs}$ are reported in Table~\ref{res:tab}. The de-lensed intrinsic effective half-light radii, R$_{\rm eff}$, have been determined by dividing $R_{\rm eff,obs}$ by the tangential magnification $\mu_{\rm tan}$ (reported in Extended Data Table~3).

\begin{table}[ht]
\caption*{{\bf Extended Data Table 1}. {\bf Cluster observed properties}. JWST photometry and uncertainties in ABmag and measured $\beta$ slopes of the candidate star clusters.} \begin{tabular}{@{}cccccccc@{}}
\toprule
ID & F150W & F200W & F277W & F356W & F410M & F444W & $\beta$\\
\midrule
A.1 & $ 27.64 \pm 0.08 $ & $ 27.83 \pm 0.07 $ & $ 27.89 \pm 0.07 $ & $ 28.13 \pm 0.09 $ & $ 27.85 \pm 0.08 $ & $ 28.03 \pm 0.10 $ & $ -2.36 ^{+ 0.19 }_{- 0.15 }$\\
B.1 & $ 27.39 \pm 0.11 $ & $ 27.48 \pm 0.08 $ & $ 27.66 \pm 0.09 $ & $ 27.83 \pm 0.10 $ & $ 28.07 \pm 0.19 $ & $ 27.90 \pm 0.10 $ & $ -2.42 ^{+ 0.23 }_{- 0.19 }$\\
C.1 & $ 27.81 \pm 0.18 $ & $ 27.78 \pm 0.17 $ & $ 27.74 \pm 0.10 $ & $ 27.86 \pm 0.10 $ & $ 27.84 \pm 0.16 $ & $ 27.94 \pm 0.11 $ & $ -1.90 ^{+ 0.27 }_{- 0.34 }$\\
D.1 & $ 27.83 \pm 0.19 $ & $ 28.02 \pm 0.19 $ & $ 27.76 \pm 0.11 $ & $ 28.05 \pm 0.13 $ & $ 27.80 \pm 0.15 $ & $ 28.06 \pm 0.12 $ & $ -1.80 ^{+ 0.35 }_{- 0.23 }$\\
E.1 & $ 28.29 \pm 0.28 $ & $ 28.39 \pm 0.22 $ & $ 28.21 \pm 0.19 $ & $ 28.66 \pm 0.24 $ & $ 28.62 \pm 0.33 $ & $ 28.24 \pm 0.16 $ & $ -1.82 ^{+ 0.43 }_{- 0.46 }$\\
A.2 & $ 28.08 \pm 0.12 $ & $ 28.07 \pm 0.09 $ & $ 28.32 \pm 0.15 $ & $ 28.87 \pm 0.21 $ & $ 28.41 \pm 0.19 $ & $ 28.60 \pm 0.23 $ & $ -2.35 ^{+ 0.37 }_{- 0.27 }$\\
B.2 & $ 27.21 \pm 0.27 $ & $ 27.19 \pm 0.23 $ & $ 27.10 \pm 0.10 $ & $ 27.33 \pm 0.10 $ & $ 27.16 \pm 0.12 $ & $ 27.45 \pm 0.14 $ & $ -1.80 ^{+ 0.31 }_{- 0.36 }$\\
C.2+D.2 & $ 27.03 \pm 0.34 $ & $ 27.08 \pm 0.20 $ & $ 27.28 \pm 0.13 $ & $ 27.45 \pm 0.13 $ & $ 27.52 \pm 0.17 $ & $ 27.21 \pm 0.14 $ & $ -2.45 ^{+ 0.46 }_{- 0.55 }$\\
\botrule
\end{tabular}
\end{table}

For each star cluster, the flux in the reference filter has been determined by integrating the fitted shape and subtracting the local background. We then measured the fluxes in the other bands by convolving the derived intrinsic shape in F150W with the empirical PSF of the respective bands.  Fitting this model to the other bands, we let the centre, normalisation, and local background (including the galaxy diffuse light) as free parameters. 

The intrinsic sizes and observed fluxes in the reference filter F150W were derived using a cutout box centered on the source with size of $11\times11$ px (about 4 times the FWHM) for A.1, A.2. Larger box sizes did not produce noticeable differences in the output sizes and fluxes. Due to their proximity, B.1, C.1, D.1, and E.1 have been fitted simultaneously within a box of $15\times15$ px. A larger box size, produces consistent values within the uncertanties of measurements for B.1 and C.1, while E.1 gets increasingly elongated, affecting the fit of D.1. To avoid this degeneracy, we fix the box size to $15\times15$, which would correspond to fix the source ellepticity of the faint E.1 to 2. Similarly, B.2,  C.2, and D.2 were fitted simultaneously within a box of $11\times11$ (changing the box size does not produce noticeable effects on the recovered parameters). We notice that we do not see two maxima at the location of C.2 and D.2, so we allowed the fit to optimise the centre of a second hidden source. We repeated the fit of this region by assuming only one source. Both approaches produced similar residuals. The flux extracted by assuming only one source is comparable within uncertainties to the flux extracted by fitting for C.2 and D.2. Due to the degeneration in identifying the position of D.2, we extracted physical properties by fitting only one source which we refer to as C.2$+$D.2. Due to the faintness of E.2 (2$\sigma$), our method did not produce significant constraints. We therefore excluded E.2 from our analysis. In the \textsc{BAGPIPES-}\emph{exp} fit we find that C.1 and D.1 have similar ages and a combined total mass of about $2\times10^6$ \msun. C.2$+$D.2 has a slightly older age (but in agreement within 1$\sigma$) than C.1 and D.1. The total mass and size of C.2$+$D.2 is a factor of two higher than their counterparts, corroborating the idea that the two star clusters are blended in \emph{Img.2}.

The size uncertainties were derived by bootstrapping the fit of the source taking into account the RMS of the local background. The photometric errors include the latter uncertainties as well as the sum in quadrature of the local background variance estimated within the box where the sources have been fitted. Aperture corrections have been extrapolated up to 0.4\arcsec\ in all bands. 

\begin{figure}
    \centering
    \includegraphics[width=1\textwidth]{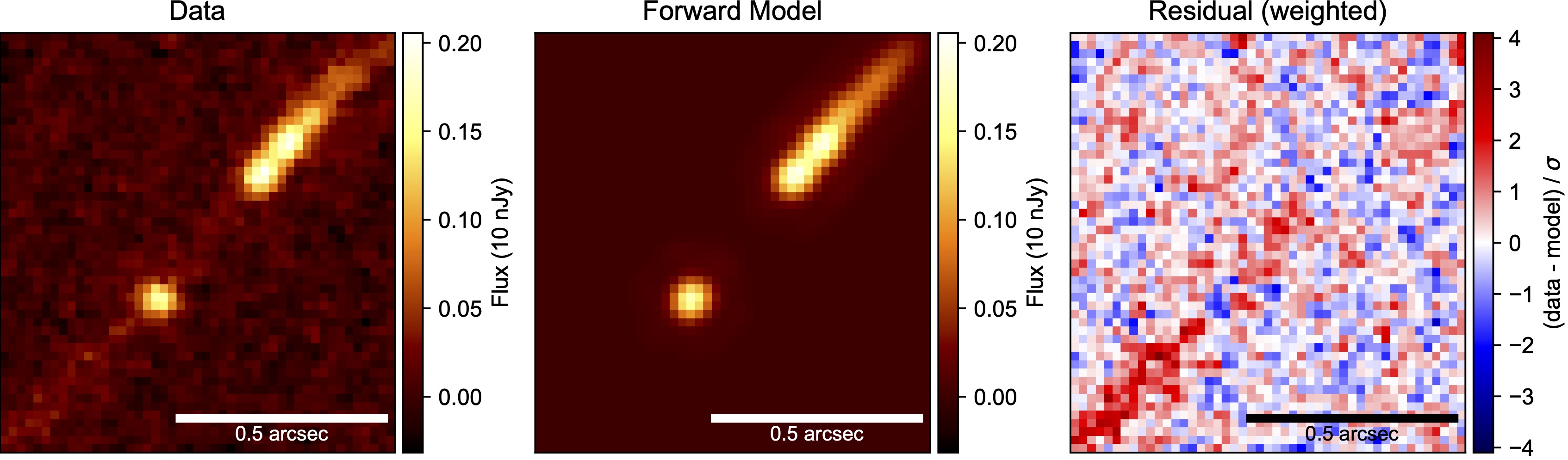}
 \caption*{{\bf Extended Data Figure 2}. {\bf Cluster light fit with forward modeling from the source plane}. Observed JWST image of \emph{Img.1} is shown (far left) along with the best fit image plane from forward modeling (left center) and weighted fit residuals (right). Weighted residuals are calculated as (data -- model)/(uncertainty).} 
\end{figure}

In Extended Data Fig.~1, we show the best model of the star clusters and the residual image in the reference filter and two more bands. The extraction of the sources does not produce significant artificial residuals above the RMS of the image. 

Independent measurements of intrinsic R$_{\rm eff}$ have been obtained following the forward modeling method of \cite{welch2023}. Briefly, this method creates a model of the galaxy in the source plane, then projects that model into the image plane. After convolving with the measured empirical PSF, the image plane model is compared to the observed data. The source plane model parameters are first optimized using a downhill simplex algorithm, then sampled using an MCMC with the Python package \texttt{emcee} \citep{foremanmackey2013}. 

For SPT0615-JD1, the source plane model consisted of five Sersic profiles centered on the five identified clumps A.1--E.1 (Extended Data Fig.~2). No diffuse component of the arc has been included in this analysis, due to the faintness of this component with respect to the clusters. Separately, we modelled clumps A.2--C.2 on the other side of the lensing critical curve. Uncertainties in the \emph{Lenstool-A} lens model resulted in slight offsets between source plane positions of clumps on opposite sides of the critical curve, which prevents simultaneous fitting of the two images of the arc. We found similar results for clump sizes on both sides of the critical curve, with clump radii ranging from 0.7--1.1 pc (see Table~1).

\subsection*{SED fitting analysis}
We performed SED fitting with \textsc{BAGPIPES} \citep{carnall2018} and test the derived physical properties against different assumptions, as well as with a different software \textsc{PROSPECTOR} \citep{johnson2021}. 
For all the runs we fix the redshift at $z=10.2$, as measured by \citep{bradley2024}. The standard stellar population templates were reprocessed with \textsc{Cloudy} to generate nebular continuum and line information (see \citep{Pacifici2023} for comparisons of the two code implementations). In both codes we assume a Kroupa IMF, unless otherwise specified. We constrain SFHs to prescriptions that reproduce a short burst in all tests except one where $\tau$ is free to vary. The short burst assumption is in agreement with studies of stellar cluster and GC populations in the local universe (\citep{adamo2020, gratton2019}). The recovered median of the posterior distributions of age, mass, A$_V$, metallicity, and associated 68\% uncertainties are reported in Extended Data Table~2.
We let the ionization parameter, $U$, to change between $-2$ and $-3.5$. We assumed a Calzetti attenuation \cite{calzetti2000} but test also the SMC extinction. In the reference set, used to produce results reported in Fig.~2 and Table~1 and referred to as \textsc{BAGPIPES-}\emph{exp}, we assumed an exponential decline with a very short $\tau=1$ Myr and Calzetti attenuation. The panels in Extended Data Fig.~3 show the observed SEDs of the 5 star clusters identified in \emph{Img.1} (black dots with uncertainties). When available, we include the observed SED of the corresponding clusters in \emph{Img.2} (orange stars with associated errors). The latter have been normalized by the median flux ratio in the 6 bands of the corresponding source in \emph{Img.1} to match the flux level while preserving the intrinsic SED shape. The best spectral and integrated photometry model obtained for the \emph{BAGPIPES-exp} fit is included. The overall shape of the observed SEDs of mirrored clusters in both images are similar within uncertainties, confirming the symmetry.  

\begin{figure}[ht]%
\centering
\includegraphics[width=1\textwidth]{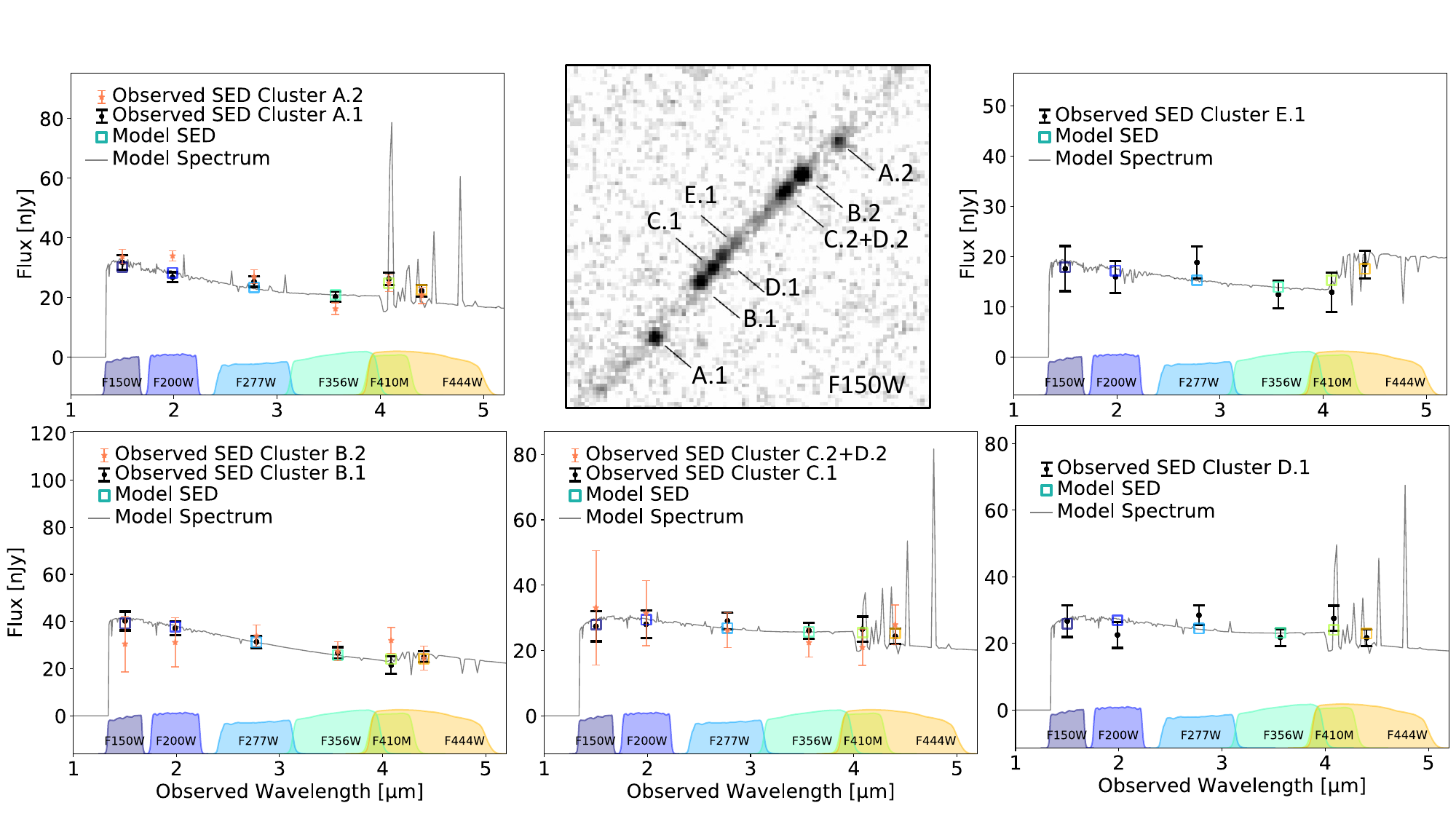}
\caption*{{\bf Extended Data Figure 3}.  {\bf 
Observed photometry and spectral energy distributions (SEDs) of each star cluster (presented in Methods).} We include the observed SEDs of the mirrored image of clusters A, B, C (orange symbols) normalised by the median ratio of the 6 bands, preserving the SED shape.} 

\end{figure}

To check the consistency of the derived cluster physical properties, we test different assumptions. The outputs are summarised in Extended Data Table~2 where we list the median and 68\% values produced by the different fits. Changing the attenuation prescription from Calzetti to SMC, produces noticeable smaller A$_v$, but all the recovered parameters are still within the 68\% uncertanties associated with the recovered values.  In  \textsc{BAGPIPES-}\emph{burst}, we assumed a single burst. We recovered slightly older ages and larger masses (but notice uncertanties) which would prefer higher stellar surface densities and older dynamical ages, confirming that we are looking at dense and bound star clusters. In a third SED fitting set, \textsc{BAGPIPES-}\emph{BPASS}, we used BPASS v2.2.1 SED templates \citep{BPASS} and the fiducial BPASS IMF with maximum stellar mass of 300 \msun and a high-mass slope similar to \cite{Kroupa2001}. Also, this model reproduces values which are very close to the reference one, suggesting that the clusters are compatible with being slightly older and therefore less sensitive to presence of very massive stars/binary systems in their light (and the limitation of fitting only 6 broad/medium band covering FUV-blue optical). Letting $\tau=free$ (we report mass-weighted parameters in Extended Data Table~2) produces significantly older ages, but similar masses, thus not affecting the results presented in this article.

\begin{sidewaystable}[ht]
\begin{tabular*}{\textwidth}{@{\extracolsep\fill}lccccccccc}
\toprule%
& \multicolumn{4}{@{}c@{}}{BAGPIPES $z=10.2$, $\tau=$1 Myr, Calzetti attenuation} & \multicolumn{5}{@{}c@{}}{Prospector SSP, Calzetti attenuation} \\
\midrule
ID & Age [Myr] &  $10^6$M$_{*}$ [\msun]& A$_V$ [mag] & Z/Z$_{\odot}$ [\%]& Age [Myr] &  $10^6$M$_{*}$ [\msun]& A$_V$[mag]  & Z/Z$_{\odot}$ [\%]&\\
\midrule
A.1 & $ 9.2 ^{+ 13.5 }_{- 3.2 }$&$ 138.34 ^{+ 293.88 }_{- 87.99 }$&$ 0.22 ^{+ 0.15 }_{- 0.14 }$&$ 6.38 ^{+ 2.42 }_{- 3.08 }$&$ 4.0 ^{+ 9.5 }_{- 0.0 }$&$ 21.94 ^{+ 328.83 }_{- 0.28 }$&$ 0.00 ^{+ 0.27 }_{- 0.00 }$&$ 16.91 ^{+ 0.87 }_{- 11.14 }$\\
B.1 & $ 14.0 ^{+ 6.5 }_{- 4.2 }$&$ 289.37 ^{+ 118.86 }_{- 137.65 }$&$ 0.20 ^{+ 0.12 }_{- 0.12 }$&$ 3.42 ^{+ 4.02 }_{- 2.49 }$&$ 51.8 ^{+ 8.6 }_{- 21.4 }$&$ 920.51 ^{+ 57.75 }_{- 238.96 }$&$ 0.04 ^{+ 0.19 }_{- 0.03 }$&$ 0.83 ^{+ 0.67 }_{- 0.34 }$\\
C.1 & $ 9.1 ^{+ 9.2 }_{- 2.8 }$&$ 172.51 ^{+ 271.70 }_{- 100.24 }$&$ 0.36 ^{+ 0.09 }_{- 0.16 }$&$ 4.56 ^{+ 3.49 }_{- 2.88 }$&$ 14.9 ^{+ 23.3 }_{- 4.7 }$&$ 585.77 ^{+ 268.97 }_{- 262.10 }$&$ 0.49 ^{+ 0.08 }_{- 0.29 }$&$ 0.57 ^{+ 1.82 }_{- 0.41 }$\\
D.1 & $ 10.9 ^{+ 13.1 }_{- 3.9 }$&$ 235.49 ^{+ 256.24 }_{- 154.93 }$&$ 0.32 ^{+ 0.12 }_{- 0.16 }$&$ 5.27 ^{+ 3.12 }_{- 3.03 }$&$ 16.2 ^{+ 19.6 }_{- 5.6 }$&$ 528.51 ^{+ 327.14 }_{- 168.52 }$&$ 0.51 ^{+ 0.07 }_{- 0.27 }$&$ 0.45 ^{+ 2.17 }_{- 0.29 }$\\
E.1 & $ 36.8 ^{+ 19.7 }_{- 16.4 }$&$ 421.73 ^{+ 153.99 }_{- 151.94 }$&$ 0.21 ^{+ 0.18 }_{- 0.15 }$&$ 5.05 ^{+ 3.15 }_{- 3.14 }$&$ 51.8 ^{+ 29.9 }_{- 24.2 }$&$ 582.23 ^{+ 254.82 }_{- 266.12 }$&$ 0.11 ^{+ 0.27 }_{- 0.09 }$&$ 2.66 ^{+ 9.38 }_{- 2.08 }$\\
A.2 & $ 17.2 ^{+ 8.8 }_{- 6.2 }$&$ 166.70 ^{+ 90.28 }_{- 77.77 }$&$ 0.11 ^{+ 0.11 }_{- 0.07 }$&$ 4.39 ^{+ 3.86 }_{- 3.28 }$&$ 8.0 ^{+ 18.7 }_{- 0.1 }$&$ 45.26 ^{+ 278.14 }_{- 1.60 }$&$ 0.01 ^{+ 0.23 }_{- 0.01 }$&$ 19.27 ^{+ 0.67 }_{- 15.80 }$\\
B.2 & $ 9.3 ^{+ 6.0 }_{- 2.6 }$&$ 294.46 ^{+ 314.14 }_{- 157.81 }$&$ 0.28 ^{+ 0.13 }_{- 0.17 }$&$ 4.83 ^{+ 3.41 }_{- 2.89 }$&$ 13.6 ^{+ 12.1 }_{- 3.1 }$&$ 743.34 ^{+ 181.02 }_{- 256.28 }$&$ 0.40 ^{+ 0.15 }_{- 0.27 }$&$ 1.24 ^{+ 3.65 }_{- 1.02 }$\\
C.2+D.2 & $ 18.6 ^{+ 7.8 }_{- 7.4 }$&$ 603.09 ^{+ 135.33 }_{- 263.11 }$&$ 0.19 ^{+ 0.16 }_{- 0.13 }$&$ 6.29 ^{+ 2.63 }_{- 3.50 }$&$ 24.5 ^{+ 11.0 }_{- 11.0 }$&$ 794.45 ^{+ 136.02 }_{- 308.30 }$&$ 0.13 ^{+ 0.27 }_{- 0.11 }$&$ 3.71 ^{+ 6.39 }_{- 3.25 }$\\
\midrule
& \multicolumn{4}{@{}c@{}}{BAGPIPES $z=10.2$, Burst, Calzetti attenuation} & \multicolumn{5}{@{}c@{}}{BAGPIPES $z=10.2$, BPASS, $\tau=$1 Myr, Calzetti attenuation} \\
\midrule
A.1 & $ 23.5 ^{+ 4.2 }_{- 17.0 }$&$ 503.90 ^{+ 106.02 }_{- 409.27 }$&$ 0.32 ^{+ 0.09 }_{- 0.17 }$&$ 0.74 ^{+ 4.93 }_{- 0.70 }$&$ 16.68 ^{+ 10.41 }_{- 7.65 }$&$ 257.14 ^{+ 159.40 }_{- 171.21 }$&$ 0.09 ^{+ 0.10 }_{- 0.06 }$&$ 3.33 ^{+ 3.66 }_{- 2.19 }$&\\
B.1 & $ 15.4 ^{+ 6.0 }_{- 6.6 }$&$ 421.94 ^{+ 129.62 }_{- 229.56 }$&$ 0.31 ^{+ 0.10 }_{- 0.17 }$&$ 0.17 ^{+ 0.75 }_{- 0.14 }$&$ 18.05 ^{+ 8.39 }_{- 5.67 }$&$ 319.09 ^{+ 153.12 }_{- 133.81 }$&$ 0.09 ^{+ 0.08 }_{- 0.06 }$&$ 0.63 ^{+ 0.90 }_{- 0.42 }$&\\
C.1 & $ 7.0 ^{+ 13.8 }_{- 4.7 }$&$ 149.70 ^{+ 434.40 }_{- 99.12 }$&$ 0.37 ^{+ 0.09 }_{- 0.20 }$&$ 0.30 ^{+ 2.71 }_{- 0.27 }$&$ 14.32 ^{+ 14.20 }_{- 11.11 }$&$ 269.70 ^{+ 275.99 }_{- 236.38 }$&$ 0.17 ^{+ 0.15 }_{- 0.11 }$&$ 1.69 ^{+ 2.63 }_{- 1.09 }$&\\
D.1 & $ 8.4 ^{+ 16.0 }_{- 3.7 }$&$ 173.50 ^{+ 391.91 }_{- 116.78 }$&$ 0.36 ^{+ 0.10 }_{- 0.18 }$&$ 0.57 ^{+ 3.84 }_{- 0.53 }$&$ 20.70 ^{+ 10.67 }_{- 8.72 }$&$ 373.07 ^{+ 166.15 }_{- 196.33 }$&$ 0.20 ^{+ 0.15 }_{- 0.12 }$&$ 1.94 ^{+ 3.61 }_{- 1.31 }$&\\
E.1 & $ 41.0 ^{+ 24.2 }_{- 14.4 }$&$ 439.84 ^{+ 133.47 }_{- 112.90 }$&$ 0.22 ^{+ 0.17 }_{- 0.14 }$&$ 0.22 ^{+ 2.52 }_{- 0.19 }$&$ 31.75 ^{+ 22.97 }_{- 14.31 }$&$ 324.68 ^{+ 158.77 }_{- 127.03 }$&$ 0.17 ^{+ 0.16 }_{- 0.11 }$&$ 3.66 ^{+ 4.00 }_{- 2.55 }$&\\
A.2 & $ 18.6 ^{+ 8.0 }_{- 6.3 }$&$ 222.91 ^{+ 73.55 }_{- 74.71 }$&$ 0.19 ^{+ 0.15 }_{- 0.11 }$&$ 0.28 ^{+ 2.60 }_{- 0.24 }$&$ 18.49 ^{+ 9.68 }_{- 6.37 }$&$ 171.49 ^{+ 99.97 }_{- 78.39 }$&$ 0.06 ^{+ 0.08 }_{- 0.04 }$&$ 1.33 ^{+ 2.20 }_{- 0.91 }$&\\
B.2 & $ 7.7 ^{+ 4.3 }_{- 2.3 }$&$ 263.42 ^{+ 302.06 }_{- 125.30 }$&$ 0.33 ^{+ 0.11 }_{- 0.16 }$&$ 1.04 ^{+ 3.22 }_{- 0.98 }$&$ 12.30 ^{+ 8.04 }_{- 8.31 }$&$ 382.18 ^{+ 286.08 }_{- 324.66 }$&$ 0.14 ^{+ 0.13 }_{- 0.10 }$&$ 2.04 ^{+ 3.46 }_{- 1.37 }$&\\
C.2+D2 & $ 19.9 ^{+ 7.3 }_{- 10.0 }$&$ 670.79 ^{+ 100.62 }_{- 317.92 }$&$ 0.19 ^{+ 0.17 }_{- 0.14 }$&$ 0.99 ^{+ 3.98 }_{- 0.93 }$&$ 19.18 ^{+ 8.00 }_{- 7.78 }$&$ 521.93 ^{+ 201.93 }_{- 248.80 }$&$ 0.11 ^{+ 0.13 }_{- 0.08 }$&$ 3.02 ^{+ 4.10 }_{- 1.77 }$\\

\midrule
& \multicolumn{4}{@{}c@{}}{BAGPIPES $z=10.2$, $\tau=$1 Myr, SMC extinction} & \multicolumn{5}{@{}c@{}}{BAGPIPES $z=10.2$, $\tau=$free, Calzetti attenuation } \\
\midrule
ID & Age [Myr] &  $10^6$M$_{*}$ [\msun]& A$_V$ [mag] & Z/Z$_{\odot}$ [\%]& Age [Myr] &  $10^6$M$_{*}$ [\msun]& A$_V$[mag]  & Z/Z$_{\odot}$ [\%]& $\tau$ [Myr]\\
\midrule
A.1 & $ 19.3 ^{+ 12.9 }_{- 13.4 }$&$ 278.55 ^{+ 166.88 }_{- 234.09 }$&$ 0.04 ^{+ 0.04 }_{- 0.03 }$&$ 6.67 ^{+ 2.39 }_{- 3.60 }$&$ 27.90 ^{+ 21.74 }_{- 14.82 }$&$ 192.02 ^{+ 135.05 }_{- 147.57 }$&$ 0.08 ^{+ 0.08 }_{- 0.05 }$&$ 6.39 ^{+ 2.40 }_{- 3.13 }$& 60.79 \\
B.1 & $ 16.4 ^{+ 6.3 }_{- 5.7 }$&$ 272.67 ^{+ 114.16 }_{- 115.28 }$&$ 0.05 ^{+ 0.03 }_{- 0.03 }$&$ 4.62 ^{+ 3.02 }_{- 3.17 }$&$ 19.15 ^{+ 14.53 }_{- 7.31 }$&$ 162.58 ^{+ 97.53 }_{- 5.19 }$&$ 0.04 ^{+ 0.06 }_{- 0.03 }$&$ 2.16 ^{+ 2.54 }_{- 1.55 }$& 65.22 \\
C.1 & $ 12.6 ^{+ 8.2 }_{- 4.8 }$&$ 259.80 ^{+ 160.50 }_{- 157.52 }$&$ 0.16 ^{+ 0.06 }_{- 0.07 }$&$ 4.10 ^{+ 3.46 }_{- 2.76 }$&$ 15.68 ^{+ 16.10 }_{- 9.40 }$&$ 171.97 ^{+ 137.98 }_{- 69.69 }$&$ 0.20 ^{+ 0.12 }_{- 0.11 }$&$ 3.84 ^{+ 3.36 }_{- 2.33 }$& 60.52 \\
D.1 & $ 13.6 ^{+ 10.7 }_{- 6.0 }$&$ 249.39 ^{+ 179.19 }_{- 168.07 }$&$ 0.15 ^{+ 0.07 }_{- 0.07 }$&$ 4.53 ^{+ 3.60 }_{- 3.00 }$&$ 18.39 ^{+ 22.90 }_{- 10.93 }$&$ 167.89 ^{+ 176.63 }_{- 86.57 }$&$ 0.19 ^{+ 0.13 }_{- 0.11 }$&$ 5.08 ^{+ 2.97 }_{- 3.12 }$& 69.27 \\
E.1 & $ 35.3 ^{+ 17.9 }_{- 14.6 }$&$ 357.91 ^{+ 148.91 }_{- 112.66 }$&$ 0.11 ^{+ 0.08 }_{- 0.08 }$&$ 5.38 ^{+ 3.27 }_{- 3.66 }$&$ 54.85 ^{+ 42.68 }_{- 30.97 }$&$ 314.15 ^{+ 176.71 }_{- 68.91 }$&$ 0.19 ^{+ 0.15 }_{- 0.13 }$&$ 4.15 ^{+ 3.79 }_{- 2.70 }$& 67.05 \\
A.2 & $ 18.3 ^{+ 7.8 }_{- 6.6 }$&$ 157.81 ^{+ 78.97 }_{- 69.06 }$&$ 0.04 ^{+ 0.04 }_{- 0.02 }$&$ 4.02 ^{+ 3.60 }_{- 2.76 }$&$ 27.56 ^{+ 22.23 }_{- 13.42 }$&$ 123.81 ^{+ 97.75 }_{- 35.06 }$&$ 0.05 ^{+ 0.07 }_{- 0.03 }$&$ 3.63 ^{+ 3.32 }_{- 2.34 }$& 65.64 \\
B.2 & $ 11.1 ^{+ 6.6 }_{- 3.5 }$&$ 350.70 ^{+ 271.40 }_{- 194.18 }$&$ 0.14 ^{+ 0.07 }_{- 0.07 }$&$ 4.44 ^{+ 3.83 }_{- 2.76 }$&$ 15.87 ^{+ 15.57 }_{- 7.76 }$&$ 270.99 ^{+ 209.38 }_{- 114.48 }$&$ 0.16 ^{+ 0.12 }_{- 0.10 }$&$ 4.43 ^{+ 3.55 }_{- 2.59 }$& 64.48 \\
C.2+D2 & $ 22.6 ^{+ 6.2 }_{- 7.3 }$&$ 632.09 ^{+ 120.30 }_{- 206.55 }$&$ 0.08 ^{+ 0.08 }_{- 0.06 }$&$ 5.56 ^{+ 2.86 }_{- 3.05 }$&$ 34.25 ^{+ 16.40 }_{- 19.41 }$&$ 474.27 ^{+ 173.05 }_{- 48.73 }$&$ 0.13 ^{+ 0.15 }_{- 0.09 }$&$ 3.79 ^{+ 3.77 }_{- 2.60 }$& 69.46 \\

\botrule
\end{tabular*}
\caption*{{\bf Extended Data Table 2}. {\bf Compilations of different SED fit outputs described in Methods.} The reported masses are not corrected for magnification. The output of the fit assuming $\tau=1$ Myr and Calzetti attenuation is referred to as \textsc{BAGPIPES-}\emph{exp} and used to derive the physical values reported in Table~1. For BAGPIPES fit with $\tau=$free we report mass weighted quantities.} 
\end{sidewaystable}

\textsc{PROSPECTOR} allows us to test single stellar population (SSP) SFH. In this case, we find that the age of A.1 is slightly younger (but within uncertanties) than those produced by the \textsc{BAGPIPES}, resulting in lower masses. This would result in slightly lower intrinsic mass M$_{*, int}=0.39\times10^6$ \msun, $\log(\Sigma_*) = 4.5$ \msun/pc$^2$, and $\log(\Pi)=1.2$, but leaving unchanged any of the conclusions of the article.

Finally, given that some of the knots in the Cosmic Gems arc are unresolved (C.1 and B.2) or only marginally resolved (A.1) in our current images, we have also explored scenarios in which these sources are individual, highly magnified stars. Using SED models for stars at high redshifts \cite{Zackrisson23} we find that, while the slopes of the SEDs of these sources would be broadly consistent with individual stars at effective temperatures $\gtrsim 20000$ K, such scenarios would require magnifications well in excess of what our macrolens models predict at the positions of these sources. Even the most massive and luminous stars (initial mass 560--575 $M_\odot$) described by the stellar evolutionary tracks of \cite{Szecsi22} would require magnifications $\mu>1000$ to explain the observed fluxes of C.1, B.2 or A.1.

\subsection*{Lens Models \& uncertainties on the derived star cluster physical properties}

Four different lensing models have been created for the SPT-CL~J0615$-$5746 cosmological field. The models are presented in detail in \citep{bradley2024}. We include here below a short description. 

\emph{Lenstool-A}, here used as reference model for the analysis presented in this letter, is based on the software \texttt{LENSTOOL} \citep{Jullo2007} which uses a parametric approach and MCMC sampling of the parameter space to identify the best-fit model and uncertainties.
In \emph{Lenstool-A},  we model the cluster lens as a combination of three main halos and cluster member galaxies, all parameterized as pseudo-isothermal mass distributions. The model uses as constraints the positions of 43 multiple images of 14 clumps, belonging to 9 unique source galaxies. The redshifts of three sources are used as constraints (the $z=10.2$ arc, and sources at $z=1.358$ and $z=4.013$ \cite{Paterno-mahler2018}), while the rest of the redshifts are treated as free parameters. 
Three clumps on each side of the main arc were used as constraints, A, B, and C, assumed to be at $z=10.2$. The model predicts a counter image at (R.A., Decl.)=($93.9490607$,$-57.7701814$). A possible candidate of this counter image, observed near this location ($\sim 1.8''$), was not used as constraint. The image plane rms of the best-fit model is $0.36$\arcsec.
All the observed lensed features are well-reproduced by this model. 

The second model, here referred to as \emph{Lenstool-B}, uses the same algorithm, but with noticeable different assumptions. This model uses 43 multiple images from 11 unique sources. A secondary cluster scale halo is placed around the location of dusty galaxies nearly 50 arcsec north of the bright centre galaxy and allowed to move within a 20\arcsec\ box around this position. As with the previous model, the $z=10.2$ arc has a predicted counter image near the possible candidate and is only about 2\arcsec\ away from the \emph{Lenstool-A} model. The main differences between those models are the assumptions about the mass distribution of the lens and the addition of constraints. The image plane rms of the best-fit model is 0.68\arcsec.

The third model used in this analysis has been created with {\tt glafic}. The {\tt glafic} \cite{2010PASJ...62.1017O,2021PASP..133g4504O} mass model is constructed with three elliptical NFW \cite{1997ApJ...490..493N} halos, external
shear, and cluster member galaxies modeled by pseudo-Jaffe
profile. 
The model parameters
are fitted to reproduce the position of 44 multiple images generated from
15 background sources. Spectroscopic redshifts are available for 7 of
the 15 sources. We include positions of A.1/A.2 and B.1/B.2 in
the Cosmic Gems arc as constraints, with small positional errors of
0.04\arcsec\ to accurately predict magnifications of each star cluster
images. For the other multiple images, we adopt the positional error
of $0.4$\arcsec. Our best-fitting model reproduces all the multiple image
positions with the root-mean-square of image positions of $0.41$\arcsec.

As a consistency check, we excluded the positional constraints from the
Cosmic Gems arc to construct the mass model, and confirmed that the
critical curve of this mass model still pass through the arc. 
Our {\tt glafic} best-fitting mass model also predicts a counter-image of the Cosmic Gems arc at around (R.A.,
Dec.)=($93.9504865$, $-57.7696559$). We find that there is a candidate
counter-image at $\sim 2$\arcsec\ from the predicted position, (R.A.,
Dec.)=($93.9500002$, $-57.7702197$). Both the consistency check and the presence of the candidate
counter-image confirm the validity of this mass model.

A fourth model has also been produced with {\tt WSLAP+} \citep{Diego2005, Diego2007}. The {\tt WSLAP+}  lens models 
 offer an alternative to parametric models and are free of assumptions made about the distribution of dark matter. When the $z=10.2$ arc is not included as a constraint, the {\tt WSLAP+} model predicts the critical curve passing at $\approx 1$\arcsec\ from the $z=10.2$ arc. This solution predicts a mirrored image of the arc which is not observed, reinforcing the expectation that the Cosmic Gems is a double image with the critical curve passing through the middle. When the arc is included as a constraint, the predicted critical curve passes between C.1 and D.1, just 0.3\arcsec\ from the alleged symmetry point in the arc, and within the uncertainties typical of {\tt WSLAP+} models. Also, this model predicts the position of a third counterimage consistent with the previous models. This model is currently under development with the goal of explaining the perturbation seen in \emph{Img.2}, and is therefore not included in this analysis. 

{\bf The photometric redshift of the candidate counter image is
$z_{phot} = 10.8_{-1.4}^{+0.6}$
(95\% confidence) \citep{bradley2024}, in agreement with the expectation}.

The total and tangential magnifications at the position of the star clusters, $\mu_{\rm tot}=\mu_{\rm tang}\times\mu_{\rm rad}$, are reported in Extended Data Table~3. For the two {\tt Lenstool}-based models, we estimated uncertainties following the method presented in \cite{claeyssens2023} based on magnification maps produced from the lenstool MCMC posterior distributions of the lens model. Uncertainties are omitted for the {\tt glafic} model. 

\begin{table}[ht]
\caption*{{\bf Extended Data Table 3}. {\bf Magnifications and associated uncertainties for 3 different lens models}. The model \emph{Lenstool-A} is used as reference in the analysis.} %
\begin{tabular}{@{}llllll@{}}
\toprule
ID & RA  & DEC & $\mu_{\rm Lenstool-A}$ ($\mu_{\rm tan}$) & $\mu_{\rm Lenstool-B}$ ($\mu_{\rm tan}$) & $\mu_{\rm glafic}$ ($\mu_{\rm tan}$) \\
\midrule
A.1 & 93.979828 & -57.772475 & $ 56.5 ^{+ 10.1 }_{ -8.9 }$(44.8) & $ 122.0^{+ 48.8 }_{ -17.8 } (81.9)$ &   76.9 (68.2)\\
B.1 & 93.979698 & -57.772395 & $ 109.2 ^{+ 31.2 }_{ -30.3 }$(84.8)& $ 212.6^{+ 152.8 }_{ -25.4 } (142.3 )$& 153.8  (125.7) \\
C.1 & 93.979660 & -57.772373  & $ 153.3 ^{+ 91.9 }_{ -55.8 }$(120.8)& $ 280.0^{+ 270.7 }_{ -27.2 }(187.9)$ & 200.9  (177.5) \\
D.1 & 93.979635 & -57.772355 & $ 209.1 ^{+ 132.8 }_{ -93.0 }$(159.8)& $ 349.5^{+ 474.6 }_{ -56.5 }(234.2)$ & 288.8  (254.7) \\
E.1 & 93.979599 & -57.772335  & $ 419.3 ^{+ 215.7 }_{ -261.6 }$(341.2) & $ 527.3^{+ 708.6 }_{ -30.1 } (353.7)$& 516.4  (455.7)\\
A.2 & 93.979316 & -57.772180  & $ 57.7 ^{+ 27.3 }_{ -3.7 }$(46.5)& $ 128.9^{+ 51.6 }_{ -17.1 }(89.5)$ & 76.3  (70.1) \\
B.2 & 93.979416 & -57.772235 & $ 97.8 ^{+ 63.5 }_{ -10.0 }$(72.0)& $ 234.3^{+ 188.7 }_{ -30.8 }(157)$ & 133.0  (116.9) \\
C.2+D.2 & 93.979460 & -57.772256  & $ 138.4 ^{+ 263.7 }_{ -20.4 } $ (117) & $ 372.8^{+ 527.0 }_{ -58.2 }(249.6)$ & 207.7  (182.3) \\

\botrule
\end{tabular}

\end{table}

\begin{figure}[ht]%
\centering
\includegraphics[width=1\textwidth]{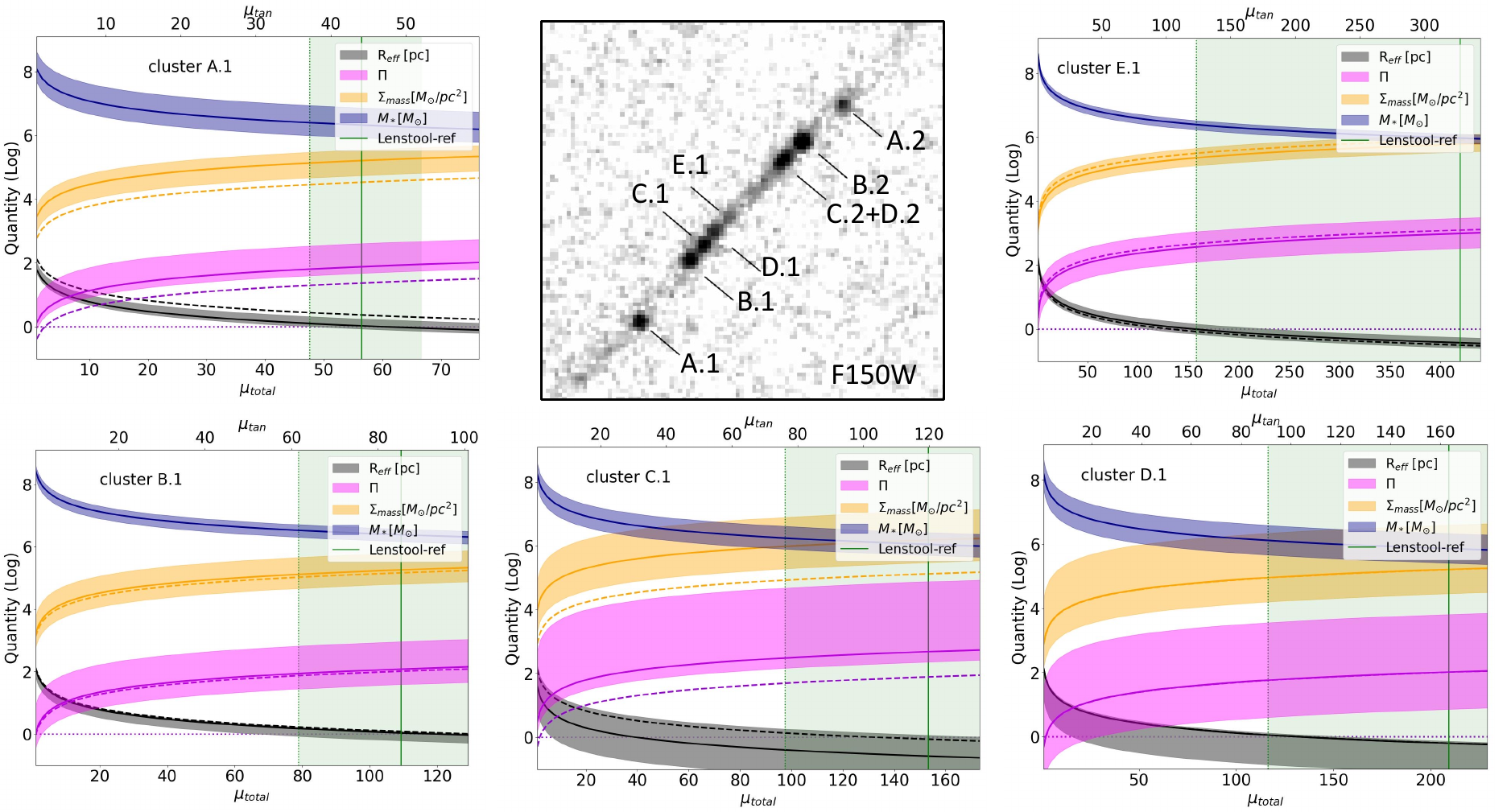}
\caption*{{\bf Extended Data Figure 4}. {\bf Measured and derived cluster physical properties as a function of their magnification.}
The most relevant quantities of each cluster in the arc (marked in the central top panel) are expressed as a function of the total magnification ($\mu_{total}$ and in case of the R$_{\rm eff}$ as a function of  $\mu_{tan}$). The radii (R$_{\rm eff}$), dynamical ages ($\Pi$), stellar mass surface densities ($\Sigma_{mass}$) and the stellar masses (M$_{\star}$) suggest the clumps are bound star clusters even at modest magnification regimes ($\mu_{total} >10$). The transparent green region and the vertical lines show the expected magnification from the reference lens model. The dotted horizontal line indicates the region where $\log(\Pi)=0$. The shaded areas in the plot mark the uncertainties associated with the derived values. The dashed lines show the lower-limits in $\Sigma_*$ and $\Pi$, assuming half the stellar PSF FWHM as upper limit for the R$_{\rm eff}$ of each star cluster.}
\end{figure}

In Extended Data Fig.~4, we show the impact that magnification predictions have in the recovered physical properties (intrinsic half-light radius and mass, black and blue solid lines) and derived quantities (dynamical age and stellar surface density in magenta and orange solid lines). We use logarithmic scales so that all quantities can be included. The colored bands show the level of uncertainties recovered from the analysis. We also include the upper-limits on the R$_{\rm eff}$ (assuming that the source is unresolved and has size smaller than the stellar PSF) and what type of the lower-limits it will translate for the physical quantities that depend on size estimates as dashed lines (notice that these are the reference quantities for C.1 which is unresolved). The magnifications (total in the bottom and tangential in the upper) are reported on the x axis. As we move to lower magnifications, the derived masses and radii become larger, consequently predicting lower stellar surface densities and dynamical ages. However, even in the unlikely case that the magnifications are wrong by one order of magnitude, stellar surface density will remain above $10^4$ \msun/pc$^2$ and dynamical ages will still be significantly larger than 1 ($\log(\Pi)>0$), leaving the main conclusion of this analysis unchanged, i.e., that we are detecting bound proto-GCs within the first 500 Myr of our Universe.

\section*{Data availability} 
The data have been acquired under JWST Program ID 4212, with PI: Bradley. The datasets generated during and/or analysed during the current study may be obtained from the MAST archive at
\url{https://dx.doi.org/10.17909/tcje-1780}. All data generated or analysed during this study are included in this published article (and its supplementary information files).

\section*{Code availability} 
This work made use of \textsc{numpy} \citep{numpy}, \textsc{scipy} \citep{scipy}, \textsc{matplotlib} \citep{matplotlib} and \textsc{astropy} \citep{astropy}. SED fit analyses are performed with publicly available software BAGPIPES \citep{carnall2018} and PROSPECTOR \citep{johnson2021}. 

\section*{Acknowledgments}

AA, EV, MM, AC, and MR thank ISSI for sponsoring the ISSI team: "Star Formation within rapidly evolving galaxies" where many ideas discussed in this article have been brainstormed. AA thanks Mark Gieles and Ivan Cabrera-Ziri for discussions on GC physical properties.
Based on observations with the NASA/ESA/CSA JWST obtained from the Mikulski Archive for Space Telescopes
(MAST) at the Space Telescope Science Institute (STScI), which
is operated by the Association of Universities for Research in
Astronomy (AURA), Incorporated, under NASA contract NAS5-03127.
Support for Program number JWST-GO-04212 was provided through
a grant from the STScI under NASA contract NAS5-03127. AA and AC acknowledge support by the Swedish research council Vetenskapsr{\aa}det (2021-05559).
J.M.D. acknowledges the support of project PID2022-138896NB-C51 (MCIU/AEI/MINECO/FEDER, UE) Ministerio de Ciencia, Investigaci\'on y Universidades. 
AZ acknowledges support by Grant No. 2020750 from the United States-Israel Binational Science Foundation (BSF) and Grant No. 2109066 from the United States National Science Foundation (NSF); by the Ministry of Science \& Technology, Israel; and by the Israel Science Foundation Grant No. 864/23. Y.J-T. acknowledges financial support from the European Union’s Horizon 2020 research and innovation programme under the Marie Skłodowska-Curie grant agreement No 898633, the MSCA IF Extensions Program of the Spanish National Research Council (CSIC), the State Agency for Research of the Spanish MCIU through the Center of Excellence Severo Ochoa award to the Instituto de Astrofísica de Andaluc\'ia (SEV-2017-0709), and grant CEX2021-001131-S funded by MCIN/AEI/ 10.13039/501100011033. EZ acknowledges project grant 2022-03804 from the Swedish Research Council (Vetenskapsr\aa{}det) and has also benefited from a sabbatical at the Swedish Collegium for Advanced Study.
M.O. acknowledges the support of JSPS KAKENHI Grant Numbers JP22H01260, JP20H05856, JP22K21349.
A.K.I. acknowledges the support of JSPS KAKENHI Grant Numbers JP23H00131. EV and MM acknowledge financial support through grants PRIN-MIUR 2020SKSTHZ, the INAF GO Grant 2022 ``The revolution is around the corner: JWST will probe globular cluster precursors and Population III stellar clusters at cosmic dawn'' and by the European Union – NextGenerationEU within PRIN 2022 project n.20229YBSAN - Globular clusters in cosmological simulations and in lensed fields: from their birth to the present epoch. 
T.H. is supported by Leading Initiative for Excellent Young Researchers, MEXT, Japan (HJH02007) and by JSPS KAKENHI grant No. 22H01258. 
Y.T.\ acknowledges the support of JSPS KAKENHI grant No.\ 22H04939 and 23K20035.

\bibliographystyle{naturemag}
\bibliography{main}

\end{document}